\documentclass{ws-ijmpe}

\newcommand{\text}[1]{{\rm {#1}}}

\newcommand{\etal}{\emph{et al.}}

\newcommand{\nuc}[2]{{$^{#1}$}{#2}}
\newcommand{\aop}{\hat{a}}
\newcommand{\aopk}{\hat{a}^\dagger}

\newcommand{\iint}{\int \! \! \! \! \int}
\renewcommand{\vec}[1]{\boldmath{#1}}

\newcommand{\tfrac}[2]{{\scriptstyle \frac{{#1}}{{#2}}}}

\newcommand{\nn}{\nonumber}

\begin{document}

\markboth{M. Bender and T. Duguet}
         {Pairing correlations beyond the mean field}

\catchline{}{}{}{}{}
%
%%%%%%%%%%%%%%%%%%%%%%%%%%%%%%%%%%%%%%%%%%%%%%%%%%%%%%%%%%%%%%%%%%%%

\title{PAIRING CORRELATIONS BEYOND THE MEAN FIELD}

\author{\footnotesize M. Bender\footnote{present 
address: CEN Bordeaux Gradignan, France}}
\address{CEA-Saclay DSM/DAPNIA/SPhN, F-91191 Gif sur Yvette Cedex, France}

\author{T. Duguet}
\address{National Superconducting Cyclotron Laboratory 
         and Department of Physics and Astronomy, 
         Michigan State University, East Lansing, MI 48824,
         USA}

\maketitle

\begin{history}
\received{(received date)}
%\revised{(revised date)}
%\accepted{(Day Month Year)}
%\comby{(xxxxxxxxxx)}
\end{history}

\begin{abstract}
We discuss dynamical pairing correlations in the context of
configuration mixing of projected self-consistent mean-field states,
and the origin of a divergence that might appear when such 
calculations are done using an energy  functional in the spirit
of a naive generalized density functional theory.
\end{abstract}
%
%==========================================================
%
\section{Introduction}
\label{sec:bmf}
Self-consistent mean-field models are one of the standard approaches 
in nuclear structure theory, see Ref.\cite{RMP} for a recent
review. For heavy nuclei, they are the only microscopic method 
that can be systematically applied on a large scale. 

Over the last few years, we were involved in the development of a
method, that adds long-range correlations to self-consistent mean-field
by projection after variation and variational configuration mixing 
within the generator coordinate method (GCM). A pedagogical introduction to
our method is given in Ref.\cite{Ben05b}, we will sketch some of our 
recent work on those aspects of this method that concern the treatment of 
pairing correlations. 
%
%==========================================================
%
\section{Theoretical framework}
%
%-----------------------------------------------------------------------
%
\subsection{Self-consistent mean-field with pairing}
For an introduction into the treatment of pairing correlations in a HFB 
framework, see for example\cite{Man75a,Rin80a,Dob84a} and references given 
therein. We will give only those details that are relevant for the further 
discussion.

All results given below were obtained with the effective Skyrme interaction 
SLy4\cite{SLyx} and a density-dependent pairing interaction with a soft
cutoff at 5 MeV above and below the Fermi energy as described in 
Ref.\cite{Rig99a}. The HFB equations are solved using the two-basis
method introduced in Ref.\cite{Gal94a}. All we need to know about an
HFB state in what follows is that in its canonical basis it is given by
\begin{equation}
| \text{HFB} \rangle
= \prod_{\mu > 0} ( u_\mu + v_\mu \aopk_{\bar\mu} \aopk_{\mu} ) 
  | 0  \rangle
.
\end{equation}
in terms of occupation probabilities % (\mbox{$u_\mu^2 + v_\mu^2 = 1$})
and creation and annihilation operators.

HFB states are not eigenstates of the particle-number operator.
In condensed matter, for which HFB theory was originally designed, 
this is not much of a problem, as the number of particles is 
usually huge. In nuclear physics, where the particle number is
quite small, this causes two problems: On the one hand, the standard 
HFB treatment artificially breaks down when the density of single-particle 
levels around the Fermi energy is below a critical value, leading to
a HF state without pairing correlations. On the other hand,
a HFB state mixes the wave functions of different nuclei as
it is spread in particle-number space, with a width that is 
proportional to the dispersion of the particle number\cite{Flo97a}. 
%
%----------------------------------------------------------
%
\section{Particle-number projection}
\label{subsec:proj}
variation after projection (VAP)\cite{She00a} provides a rigorous solution to 
both problems. For various mainly technical reasons, however,
this approach is not yet widely used (see\cite{Sto06a} for an exception). 
We use a different approach instead, that was first outlined in
Ref.\cite{Hee93a}. In a first step,
we complement the HFB equation with the Lipkin-Nogami (LN) method, that
provides a numerically simple approximation to projection before
variation, see Ref.\cite{Gal94a} and references given therein for details. 
The LN procedure enforces the presence of pairing
correlations also in the weak pairing regime at small
level density. The LN method gives a correction to the total energy 
whose quality has been repeatedly questioned. We do not make use of the
correction term, but project in a second step on exact eigenstates 
of the particle-number operator $\hat{N}$, with eigenvalue $N_0$, 
applying the particle-number projection operator
\begin{equation}
\label{eq:proj:number}
\hat{P}_{N_0}
= \frac{1}{2 \pi}
  \int_{0}^{2 \pi} \! d \varphi \;
  e^{i \varphi (\hat{N}-N_0)}
.
\end{equation}
The operator $e^{i \varphi \hat{N}}$ rotates the HFB state in a $U(1)$ 
gauge space
\begin{equation}
\label{eq:HFB:rot}
| \text{HFB} (\varphi) \rangle
= e^{i \varphi \hat{N}} | \text{HFB} \rangle
= \prod_{\mu > 0} ( u_\mu + v_\mu e^{2i\varphi} \aopk_{\bar\mu} \aopk_{\mu} ) 
  | 0  \rangle
,
\end{equation}
while $e^{-i \varphi N_0}$ is a weight function. 
%A projected HFB state
%with eigenvalue $N_0$ and normalized to 1 is given by
%%
%\begin{equation}
%| N_0 \rangle
%= \frac{\hat{P}_{N_0} | \text{HFB} \rangle}
%       {\langle \text{HFB} | \hat{P}_{N_0} | \text{HFB} \rangle^{1/2}}
%.
%\end{equation}
%%
For states with even particle number $N_0$ the integration interval
can be reduced to $[0,\pi]$. We discretize the integrals over the gauge 
angle with a simple $L$-point trapezoidal formula
\begin{equation}
\label{eq:fomenko}
\frac{1}{\pi} \int_{0}^{\pi} d{\varphi} \, e^{i\varphi(\hat{N}-N_0)}
\Rightarrow \frac{1}{L} \sum_{l=1}^{L} e^{i \frac{\pi l}{L}(\hat{N}-N_0)}
.
\end{equation}
As was shown by Fomenko\cite{Fom70a}, this simple scheme
eliminates exactly all components from an HFB state which 
differ from the desired particle number $N_0$ by up to $\pm 2(L-1)$
particles. Although the spread of the near-Gaussian distribution of
particle numbers is large in comparison with respect to the constrained 
value, it is small enough that already small values of $L$, ranging 
from 5 in light nuclei to 13 in heavy ones, are sufficient for the numerical 
convergence of the integrals over $\varphi$.

In practice, we constrain the HFB states in the mean-field calculations
to the same (integer) particle number that we project on afterwards. This is,
however, not a necessary condition. In a projection-before-variation
approach, the particle number of the intrinsic state has no physical
meaning, and will usually take a non-integer value close to, but not
identical with, the particle number projected on\cite{Flo97a,She00a}.

We address here only HFB states with pairing correlations among
particles of the same isospin. In this case, the nuclear HFB state 
is the direct product of separate HFB states for protons and 
neutrons, respectively, which are separately projected afterwards
on the number of the respective particle species.
%
%------------------------------------------------------------
%
\subsection{Variational configuration mixing}
\label{subsec:gcm} 
In a mean-field calculation, one often encounters a 
situation where the total binding energy (mean field or projected)
varies only slowly with a collective degree of freedom. In such 
a case, it can be expected that the  nuclear wave function 
is widely spread around the mean-field minimum, which is beyond 
the scope of (projected) mean-field theory.
These fluctuations around a single state can be incorporated
within the generator coordinate method (GCM). The mixed projected 
many-body state is set-up as a coherent superposition of projected 
mean-field states $| q \rangle$ which differ in one or several
collective coordinates $q$
\begin{equation}
| k \rangle
= \sum_{q} f_{k} (q) \, | q \rangle
.
\end{equation}
The weight function $f_{k} (q)$ is determined from the 
stationarity of the GCM ground state, which leads to the 
Hill-Wheeler-Griffin equation\cite{Hil53a}
\begin{equation}
\label{eq:HWG}
\sum_{q'}
\big[ \langle q | \hat{H} | q' \rangle 
      - E_k \, \langle q | q' \rangle \big] \; f_{k} (q') 
= 0
,
\end{equation}
that gives a correlated ground state, and, in addition, a spectrum of excited 
states from orthogonalization to the ground state. The weight functions 
$f_{k} (q)$ are not orthonormal. A set of orthonormal collective wave 
functions in the basis of the states $| q \rangle$ is obtained from 
a transformation involving the square root of the norm kernel\cite{Bon90a}.
It has to be noted that projection is in fact a special case of the GCM, 
where degenerate states are mixed. The generators of the group involved 
define the collective path, and the weight functions are determined 
by the restored symmetry. 

For a state which would result from the mixing of different unprojected 
mean-field states, the mean particle number will not not be equal to 
the particle number of the original mean-field states anymore. 
Projection on particle-number, 
as done here, eliminates this problem, otherwise a constraint on 
the particle number has to be added to the Hill-Wheeler-Griffin 
equation (\ref{eq:HWG}), see, for example, Ref.\cite{Bon90a}.

The technical challenges of a configuration-mixing calculation 
come from the non-diagonal kernels of different 
mean-field states, which are evaluated with a generalized 
Wick theorem\cite{Oni66a,Bal69a}. We represent the single-particle 
states on a 3-dimensional mesh in coordinate space using a 
Lagrange mesh technique\cite{Bay86a}. As a consequence, the two 
sets of single-particle states representing the intrinsic HFB states
entering the kernels are usually not equivalent, which has to be 
carefully taken into account\cite{Bon90a,Val00a}.

We usually combine the techniques presented above with a projection
on angular momentum\cite{Val00a} that will not be discussed here.
Applications are published 
in Refs.\cite{Ben03a,Ben03b,Dug03a,Ben04b,Ben04c,Ben05a,Ben06a}. 
Similar methods, but without particle-number projection, have been 
set up using the Gogny force\cite{Madrid} and relativistic 
Lagrangians\cite{Nik06a}.

For the sake of simple notation, we have introduced the GCM using 
a many-body Hamiltonian $\hat{H}$. All the methods just mentioned,
however, have in common that they are not based on a Hamiltonian, 
but an energy functional to calculate the binding energy. 
The necessary generalization will be sketched
in section \ref{sect:divergence} below.
%
%==========================================================
%
\section{Dynamical pairing correlations}
\label{sect:dyn}
There is fundamental problem with projection after variation as 
performed here: it cannot be expected that the projection of the
mean-field ground state (after variation) gives the minimum of the energy 
hyper-surface that is obtained by the projection of all possible
mean-field states (which would be found by projection before variation).
When projecting deformed mean-field states on angular-momentum, the
intrinsic deformation of the mean-field ground state will indeed 
usually be different from the intrinsic deformation of the state giving 
the minimum of the projected energy curve, with the the exception 
of well-deformed heavy nuclei in the rare-earth, actinide, and 
transactinide regions.

In the context of angular-momentum projection, we overcome this
problem to some extend by a minimization after projection (MAP), where 
we generate an energy curve of projected mean-field states with
different intrinsic quadrupole deformation, whose minimum provides a 
first order approximation to projection before variation.
When performing a GCM calculation of projected states, the spacing
of points along this deformation energy curve does not need to be very
dense, it even does not have to contain the actual minimum, as the
variational projected GCM calculation of two (non-orthogonal) 
projected states around a minimum has the ability to (implicitely) 
construct this minimum, as the projected state representing the minimum 
has a non-zero overlap with the actually used states. As the 
projected GCM ground state also describes the energy gain from 
fluctuations around this minimum, its energy will be even below that 
of the minimum of the projected energy curve. It has to be stressed 
that the
correlations from fluctuations around the projected state are outside 
the scope of projection before variation. As a consequence, projection 
before variation does not \emph{per se} give a better description of
correlations beyond the mean field than a GCM of states projected 
after variation.

While this is a standard procedure in the context of angular-momentum 
projection, where a MAP is performed with respect to quadrupole deformation, 
it has been rarely addressed in connection with particle-number projection. 
The most obvious degree of
freedom to search for a minimum of a particle-number projected 
energy curve is the amount of pairing correlations contained in the
(unprojected) intrinsic state. Using this degree of freedom in a 
projected GCM calculation is equivalent to including the ground-state 
correlations from pairing vibrations (see Refs.\cite{Rin80a}
for an overview and Refs.\cite{Rip69a,Fae73a} for 
early GCM calculations using schematic models).
Exploratory studies along these lines within the context of realistic
mean-field models are presented in Refs.\cite{Mey91a,Hee01a}. They do,
however, not involve projection on angular momentum.

We will present here a similar study of the role of dynamical pairing
correlations in \nuc{120}{Sn}. First of all, we have to note that
finding a suitable constraint on pairing correlations is not
an easy task. The obvious choice in schematic models is the pairing gap
obtained from a pairing force with constant matrix elements. This 
coordinate was, in fact, also used in Refs.\cite{Mey91a,Hee01a}. It has,
however, some serious drawbacks as this pairing force leads to unrealistic
asymptotics of the HFB state at large distances from the nucleus\cite{Dob84a}. Constraints on other observables that measure pairing correlations 
pose similar problems even when used with a 
realistic pairing interaction, or they introduce ambiguities
on how to put them into the variational equations\cite{pairconstraint}. 
The situation is similar to a constraint on a multipole moment of the 
density distribution of order $\ell$ used to generate an energy surface 
as a function of deformation. Such a constraint has always to be damped 
at large distances from the nucleus, as it introduces along some direction 
a contribution to the constrained single-particle potential that diverges as
$-r^\ell$.

%
%==============
%
\begin{figure}[t!]
\centerline{\includegraphics{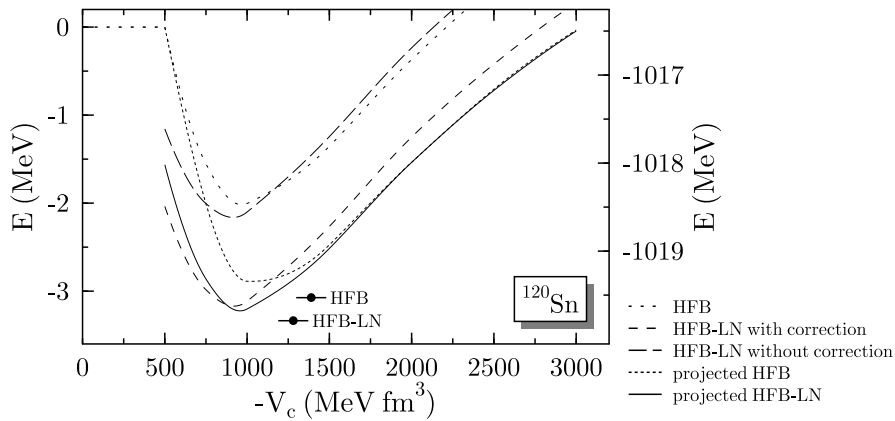}}
\caption{\label{fig:sn120}
HFB and HFB-LN mean-field energy curves, the latter with and without 
the LN correction term in the energy, and particle-number projected energy 
curves as a function of the "generating pairing strength" $V_c$ for 
a spherical mean-field state in \nuc{120}{Sn}.
The dots with horizontal bars represent the energy of the projected GCM 
ground state plotted at its average "generating pairing strength".
The energy scale on the left is normalized with respect to the HF ground 
state, while the scale to the right gives the total binding energy. 
}
\end{figure}
%
%==============
%

A not completely satisfactory, but well working constraint on the amount
of pairing correlations is provided by the strength of the pairing force. 
A calculation is done in two steps: First, the HFB or HFB-LN equations 
are solved using a "generating pairing strength" $V_c$. Then, in a second 
step, the energy of each HFB state is re-calculated without iterating 
the HFB equations using the realistic pairing strength $V_0 = -1000$ MeV 
fm$^{3}$, either with or without projection. In a third step, the projected
HFB states can be mixed within the GCM. Only the pairing strength of the
neutrons is changed. For protons, we always use $V_0 = -1000$ MeV 
fm$^{3}$, and the same method as for the decription of the neutrons;
in the case of pure HFB this means that the proton pairing breaks down.
An example of such a calculation is shown in Fig.\ \ref{fig:sn120}. 
The most remarkable findings are

1) 
The HFB equations without LN corrections break down to HF at values 
of the pairing strength around $-500$ MeV fm$^{3}$. For those states, 
our formalism cannot introduce pairing correlations beyond the mean-field.
Just above the transition from an unpaired to a paired system, the
energy gain from projection rises very rapidly up to about the $V_c$
corresponding to the minimum of the energy curve (which differes on the
order of 10 $\%$ from $V_0$), and decreases more 
slowly afterwards. This is analogue to what is found in the 
angular-momentum projection of quadrupole deformed states around a
spherical configuration.

2) 
There are obvious differences between HFB and HFB-LN energy curves, both
on the mean-field level and projected, for small generating pairing 
strength $V_c < -1500$ MeV fm$^{3}$, which is indeed in
the regime of the physical pairing strength. In the strong pairing regime 
at larger values of $V_c$, the energy curves obtained from projection 
of the HFB and HFB-LN states are nearly identical. The remaining 
difference might be attributed to the presence of proton pairing 
correlations in the HFB-LN case, while they are absent for HFB.

3) 
The LN correction overestimates the energy gain from projection 
at small generating pairing strength, and underestimates it in the
strong pairing regime. 

4)
The additional energy gain from the GCM of projected states is quite
small, around 100 keV when starting with HFB-LN states, and about
200 keV for pure HFB states. The unphysical breakdown of HFB in the weak 
pairing regime leads to 
a smaller binding energy at the minimum of the projected energy curve, 
and gives a potential energy surface that is stiffer than the HFB-LN 
one for small $V_c$. This leads to a GCM ground state from projected 
HFB that is less bound than the GCM gound state from projected HFB-LN.
This also pushes the GCM ground state wave function from projected HFB 
to larger values of $V_c$ than the one from projected HFB-LN, as seen
from the larger average generating pairing strength 
$\bar{V}_c = \sum_{V_c} g_0^2 (V_c) V_c$ in the projected HFB case.

This calculation, of course, scratches only on the surface of the
importance of dynamical pairing correlations. The question of better 
constraints on pairing correlations, systematics, excited states,
and the coupling with deformation modes will be addressed elsewhere.
%
%==========================================================
%
\section{The divergence in particle-number projection}
\label{sect:divergence}

\subsection{General features}
%
%==============
%
\begin{figure}
\centerline{\includegraphics{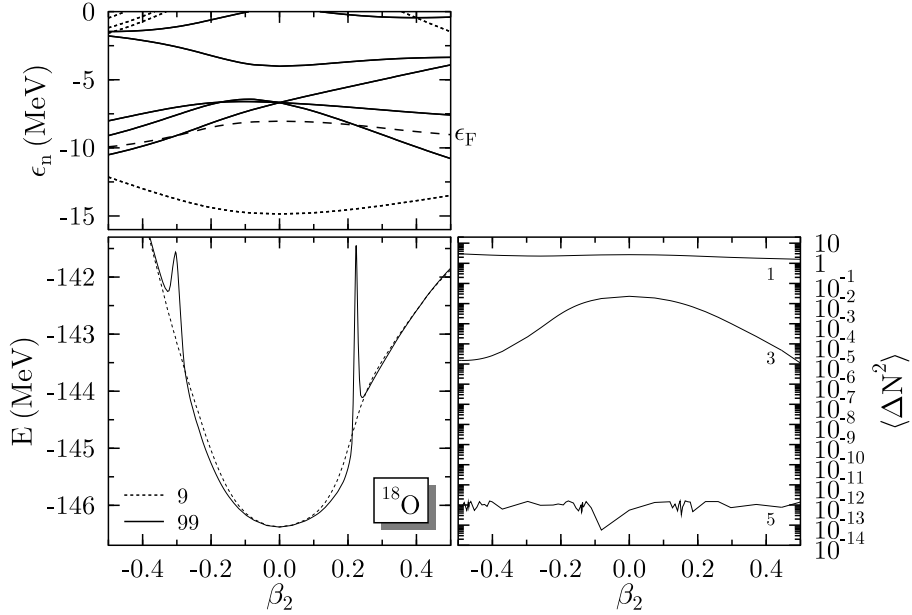}}
\caption{\label{fig:o18:sly4}
Lower panel, left: particle-number projected binding energy of \nuc{18}{O}
calculated with the Skyrme interaction Sly4 and a density-dependent
pairing interaction as a function of the mass quadrupole deformation 
for 9 and 99 discretization points for the integral over the gauge 
angle $\varphi$ in Eq.\ (\protect\ref{eq:fomenko}). Upper panel, 
left: canonical single-particle energies (full lines: parity $+1$, 
dotted lines: parity $-1$) and Fermi energy (dashed line) for neutrons.
Right panel: dispersion of the neutron number 
$\langle N_0 | \hat{N}^2 | N_0 \rangle - \langle N_0 | \hat{N} | N_0 \rangle^2$
for 1 (no projection), 3 and 5 discretization points.
}
\end{figure}
%
%==============
%
It has been noticed for a long time that the particle-number projected 
energy might exhibit divergences\cite{Don98a,Ang01b}
when a single-particle level crosses the Fermi energy and has the
occupation \mbox{$v^2_\mu = 0.5$}. More recently, it was pointed out 
by Stoitsov \etal\cite{Dob05a}, that in addition to the divergence 
there appears a finite step in the projected energy when passing this 
situation as a function of a constraint. An example is given in 
Fig.\ \ref{fig:o18:sly4}, where two clear divergences appear when
a fine discretization of the integral over gauge angles, 
Eq.\ (\protect\ref{eq:fomenko}), is used. The steps are also hinted,
with the binding energy changing on both 
sides from a lower curve around sphericity to a higher lying curve when 
passing the divergence.

First of all, it has to be stressed that the binding energy is the only 
observable that shows an anomaly when a single-particle level crosses 
the Fermi energy. Neither the overlap, nor any 
observable calculated from any $n$-body operator shows an unusual
behaviour. This is examplified in Fig.\ \ref{fig:o18:sly4} by the
dispersion of the particle number, a two-body operator that also provides 
a measure for the numerical quality of projection. For a small system
like \nuc{18}{O}, the integrals over gauge angles are already numerically 
converged for \mbox{$L=5$} discretization points. The only exception is
the binding energy. However, an extremely huge number of discretization 
points is needed to see the divergence develop in \nuc{18}{O} (and even more 
for heavier nuclei), which is one of the reasons why the divergence often 
remains undetected. The other reason is that it is very unlikely that
one of the discrete points used to calculate a potential energy 
surface hits the narrow region where the divergence 
appears. This is different when performing projection before variation,
where the variation will detect the divergences more easily\cite{Dob05a}.

The appearance of a divergence for the binding energy, but no other
observable is related to the the particular definition of the
binding energy in our method: the energy is calculated from an energy 
density functional, while everything else is, for the moment at least, 
calculated as the expectation value of an operator.

To understand the origin of the divergence, we have to look into 
the definition and evaluation of the energy functional. For the sake 
of simple notation and a transparent argument, we will restrict 
ourselves here to a toy model with one kind of particles 
only, and a two-body interaction. The additional complications 
introduced by density dependencies as needed in realistic energy 
functionals will be discussed elsewhere\cite{Ben07a}. The further 
generalization to a system composed of protons and neutrons
is then straightforward. 
All expressions given below are evaluated in the canonical 
basis, as this basis will turn out to be the only basis in which 
the origin of the divergence can be clearly identified and analyzed.
%
%-----------------------------------------------------------------------
%
\subsection{The Hamiltonian case}
As a reference, for which everything is properly defined and no problem
occurs, we will use the energy obtained from a two-body force. At the HFB
level, one has
\begin{eqnarray} 
\label{E:mf:hatH}
\mathcal{E}^{\hat{H}} [\rho, \kappa^{01}, \kappa^{10} ] 
& = & \mathcal{E}_{\text{kin}} [\rho ] 
      + \tfrac{1}{4} \sum_{ijmn \gtrless 0} \bar{v}_{ijmn} \;
        \langle \text{HFB} | \aopk_{i} \aopk_{j} \aop_{n} \aop_{m}
        | \text{HFB} \rangle 
      \\
& = & \mathcal{E}_{\text{kin}} [\rho ] 
      + \tfrac{1}{2} \sum_{\mu,\nu \gtrless 0} 
        \bar{v}_{\mu\nu\mu\nu} \, 
        \langle \aopk_{\mu} \aop_{\mu} \rangle
        \langle \aopk_{\nu} \aop_{\nu} \rangle
      + \tfrac{1}{4} \sum_{\mu,\nu \gtrless 0} \, 
        \bar{v}_{\mu\bar{\mu}\nu\bar{\nu}} \, 
        \langle \aopk_{\mu} \aopk_{\bar\mu} \rangle
        \langle \aop_{\bar\nu} \aop_{\nu} \rangle
 \nn
.
\end{eqnarray}
The contractions are the usual density matrix and pair tensor in the
canonical basis
\begin{eqnarray}
\langle \aopk_{\mu} \aop_{\nu} \rangle
& = & \rho_{\mu \nu} 
  =   \frac{\langle \text{HFB} | a^{\dagger}_{\nu} a_{\mu} 
            | \text{HFB} \rangle}
           {\langle \text{HFB} | \text{HFB} \rangle} 
  =   \rho_{\mu \mu} \, \delta_{\mu \nu} 
  =   v^{2}_{\mu} \, \delta_{\mu \nu} 
      \\
\langle \aop_{\nu} \aop_{\mu} \rangle
& = & \kappa^{01}_{\mu \nu} 
  =   \frac{\langle \text{HFB} | a_{\nu} a_{\mu} 
            | \text{HFB} \rangle}
           {\langle \text{HFB} | \text{HFB} \rangle} 
  =   \kappa^{01}_{\mu \bar{\mu}} \, \delta_{\nu\bar{\mu}}  
  =   u_{\mu} v_{\bar{\mu}} \, \delta_{\nu\bar{\mu}} 
      \\
\langle \aopk_{\mu} \aopk_{\nu} \rangle
& = & \kappa^{10}_{\mu \nu} 
  =   \frac{\langle \text{HFB} | a^{\dagger}_{\mu} a^{\dagger}_{\nu} 
            | \text{HFB} \rangle}
           {\langle \text{HFB} | \text{HFB} \rangle} 
  =   \kappa^{10}_{\mu \bar{\mu}} \, \delta_{\nu\bar{\mu}}  
\equiv u_{\mu} v_{\bar{\mu}} \, \delta_{\nu\bar{\mu}}
,
\end{eqnarray}
Unlike in papers on standard HFB theory we distinguish already here
between two different pair tensors, as they generalize differently
for particle-number projected HFB states.
Unless necessary, we will not specify the kinetic energy in what
follows. It is given by a one-body operator, always evaluated
as such from whatever the left and right many-body states might be,
and free of any divergence problems. The generalization of 
$\mathcal{E}^{\hat{H}}$ to particle-number projected states 
is straightforward
\begin{eqnarray}
\label{E:proj:hatH}
\mathcal{E}^{\hat{H}}
& = & \frac{\langle \text{HFB} | \hat{H} \hat{P}^N | \text{HFB} \rangle }
           {\langle \text{HFB} | \hat{P}^N | \text{HFB} \rangle } 
  =   \int_0^{2\pi} \! \! \frac{d \varphi}{2\pi \mathcal{D}_{N_0}} \, 
      e^{-2i\varphi N_0} \,
      \langle \text{HFB} (0) | \hat{H} | \text{HFB} (\varphi) \rangle       
\end{eqnarray}
with 
$\mathcal{D}_{N_0} = \langle \text{HFB} | \hat{P}^N | \text{HFB} \rangle$.
The relevant piece is the calculation of the Hamiltonian kernel
$\langle \text{HFB} | \hat{H} | \text{HFB} (\varphi) \rangle$
Although for particle-number projection the Hamiltonian kernel 
can -- in principle -- be evaluated using the standard Wick 
theorem, it is much more convenient to apply a generalized 
Wick theorem\cite{Oni66a,Bal69a}
\begin{eqnarray}
\label{eq:Eproj:H}
\langle \text{HFB} | \hat{H} | \text{HFB} (\varphi) \rangle  
& = & \bigg[
      \sum_{\mu \gtrless 0} t_{\mu \mu} 
      \langle \aopk_{\mu} \aop_{\mu} \rangle_\varphi
      + \tfrac{1}{2} \sum_{\mu,\nu \gtrless 0} 
        \bar{v}_{\mu\nu\mu\nu} \, 
        \langle \aopk_{\mu} \aop_{\mu} \rangle_\varphi
        \langle \aopk_{\nu} \aop_{\nu} \rangle_\varphi
      \nn \\
&   & \qquad
      + \tfrac{1}{4} \sum_{\mu,\nu \gtrless 0} \, 
        \bar{v}_{\mu\bar{\mu}\nu\bar{\nu}} \, 
        \langle \aopk_{\mu} \aopk_{\bar\mu} \rangle_\varphi
        \langle \aopk_{\bar\nu} \aopk_{\nu} \rangle_\varphi
      \bigg] \; 
      \mathcal{I} (\varphi)
. 
\end{eqnarray}
The basic contractions are given by
\begin{eqnarray}
\langle \aopk_{\mu} \aop_{\nu} \rangle_\varphi
& = & \rho_{\mu \nu} (\varphi) 
  =   \rho_{\mu \mu} (\varphi) \, \delta_{\nu \mu} 
  =   \frac{v_{\mu}^2 \, e^{2 i \varphi}}
           {u_\mu^2 + v_{\bar{\mu}}^2 \, e^{2 i\varphi} } \, \delta_{\nu \mu} 
      \\
\langle \aop_{\nu} \aop_{\mu} \rangle_\varphi
& = & \kappa^{01}_{\mu \nu} (\varphi) 
  =   \kappa^{01}_{\mu \nu} (\varphi) \, \delta_{\nu \bar{\mu}}
  =   \frac{u_\mu v_{\bar{\mu}}}
           {u_\mu^2 + v_{\bar{\mu}}^2 \, e^{2 i \varphi} } \,
      \delta_{\nu \bar{\mu}}
      \\
\langle \aopk_{\mu} \aopk_{\nu} \rangle_\varphi
& = & \kappa^{10}_{\mu \nu} (\varphi) 
  =   \kappa^{10}_{\mu \nu} (\varphi) \, \delta_{\nu \bar{\mu}}
  =   \frac{u_\mu v_{\bar{\mu}} e^{2 i \varphi}}
           {u_\mu^2 + v_{\bar{\mu}}^2 \, e^{2 i \varphi} } \,
      \delta_{\nu \bar{\mu}} 
\end{eqnarray}
and the norm kernel $\mathcal{I} (\varphi)$ by
\begin{eqnarray} 
\label{eq:I:can}
\mathcal{I} (\varphi) 
& = & \langle \text{HFB} | \text{HFB} (\varphi) \rangle  
  =   \prod_{\mu>0} (u_\mu^2 + v_\mu^2 \, e^{2 i \varphi} ) 
.
\end{eqnarray}
%
%-----------------------------------------------------------------------
%
\subsection{The EDF case}
In the case of general energy density functional (EDF) in the spirit 
of a Kohn-Sham approach with pairing, the energy is given by 
\begin{eqnarray} 
\label{E:mf:dft:1}
\mathcal{E}^{\text{EDF}} [\rho, \kappa, \kappa^{\ast}] 
& = & \mathcal{E}_{\text{kin}} [\rho ] 
      + \mathcal{E}_{\rho\rho} [\rho ] 
      + \mathcal{E}_{\kappa\kappa} [\kappa^{10}, \kappa^{01} ] 
,
\end{eqnarray}
where $\mathcal{E}_{\text{kin}}$ is the kinetic energy,
$\mathcal{E}_{\rho\rho}$ the energy of the particle-hole
interaction, and $\mathcal{E}_{\kappa\kappa}$ the energy 
from the particle-particle (pairing) interaction.
The only restriction that we impose on the energy functional 
for the moment is that it is bilinear in either the density 
matrix or the pair tensor, which is done to keep the analogy
with the Hamiltonian case. Then, the energy functional can always be 
described in terms of the kinetic energy and a double sum over 
\emph{not antisymmetrized} two-body matrix elements 
$w^{\rho\rho}_{\mu\nu\mu\nu}$ and $w^{\kappa\kappa}$, different
in the particle-hole 
\begin{eqnarray} 
\label{E:mf:dft:2}
\mathcal{E}^{\text{EDF}}
& = &   \mathcal{E}_{\text{kin}} [\rho ] 
      + \sum_{\mu\nu \gtrless 0} \, w^{\rho\rho}_{\mu\nu\mu\nu} \, 
        \rho_{\mu \mu} \, \rho_{\nu \nu}
      + \sum_{\mu\nu \gtrless 0} \, 
        w^{\kappa\kappa}_{\mu\bar{\mu}\nu\bar{\nu}} \, 
        \kappa^{10}_{\mu \bar\mu} \kappa^{01}_{\nu \bar\nu}
       \nn \\
& = &   \mathcal{E}_{\text{kin}} [\rho ] 
      + \sum_{\mu\nu \gtrless 0} \, w^{\rho\rho}_{\mu\nu\mu\nu} \, 
        \langle \aopk_{\mu} \aop_{\mu} \rangle
        \langle \aopk_{\nu} \aop_{\nu} \rangle
      + \sum_{\mu\nu \gtrless 0} \, 
        w^{\kappa\kappa}_{\mu\bar{\mu}\nu\bar{\nu}} \, 
        \langle \aopk_{\mu}    \aopk_{\bar\mu} \rangle
        \langle \aop_{\bar\nu} \aop_{\nu} \rangle
.
\end{eqnarray}
This is the Kohn-Sham approach to an EDF\cite{Koh64a}, formally generalized 
to systems with pairing by Oliveira \etal\cite{Oli88a}, where the actual 
density (matrix) is calculated from an auxiliary independent quasi-particle 
state $| \text{HFB} \rangle$.
Equation (\ref{E:mf:dft:2}) might appear to be an unusual representation 
of an energy functional, but, for example, the interaction of a term bilinear
in the local density, where $f(|\vec{r}-\vec{r}'|)$ is an arbitrary function
and $C$ the coupling constant, translates as
\begin{eqnarray}
\lefteqn{
C \iint \! d^3r \; d^3 r' \; \rho (\vec{r}) \, f(|\vec{r}-\vec{r}') \, 
                             \rho (\vec{r}')
} \nn \\
& = & C \sum_{\mu, \nu \gtrless 0} 
      \iint \! d^3r \; d^3 r' \; 
      \rho_{\mu \mu} \, \psi^\dagger_\mu (\vec{r}) \, \psi_\mu (\vec{r}) \,
      f(|\vec{r}-\vec{r}'|) \, 
      \rho_{\nu \nu} \, \psi^\dagger_\nu (\vec{r}') \, \psi_\nu (\vec{r}')
      \nn \\
& = & C \sum_{\mu, \nu \gtrless 0}
        \rho_{\mu \mu} \, w^{\rho\rho}_{\mu\nu\mu\nu} \, \rho_{\nu \nu} 
.
\end{eqnarray}
Similar expressions are obtained for any other bilinear contribution
to an energy functional that does not have density-dependent coupling
constant.

For the generalization of the energy functional to the case of
particle-number projected states, the same generalized Wick theorem 
as above is used to define
\begin{equation}
\label{eq:EDF:proj}
\mathcal{E}^{\text{EDF}} (\varphi)
= \int_0^{2\pi} \! \! \frac{d \varphi}{2\pi \mathcal{D}_{N_0}} \, 
  e^{-i\varphi N_0} \, 
  \mathcal{H}^{\text{EDF}} (\varphi)
\end{equation}
where the Hamiltonian kernels are now given by
\begin{eqnarray}
\label{eq:E:proj:dft:2}
\mathcal{H}^{\text{EDF}} (\varphi) \!
& = & \mathcal{H}_{\text{kin}} (\varphi)
      + \! \! \sum_{\mu,\nu \gtrless 0} \! \! 
        \big[
          w^{\rho\rho}_{\mu\nu\mu\nu}
          \langle \aopk_{\mu} \aop_{\mu} \rangle_\varphi
          \langle \aopk_{\nu} \aop_{\nu} \rangle_\varphi
%       \nn \\
%&   & \qquad 
%        + \sum_{\mu,\nu \gtrless 0}
         + w^{\kappa\kappa}_{\mu\bar{\mu}\nu\bar{\nu}}
          \langle \aopk_{\mu} \aopk_{\bar\mu} \rangle_\varphi
          \langle \aopk_{\bar\nu} \aopk_{\nu} \rangle_\varphi
        \big]  
      \mathcal{I} (\varphi)
\nn \\
\phantom{nothing left} 
\end{eqnarray}
The identification of the origin of the divergence proceeds 
as follows

1)
As analyzed in detail by Anguiano \etal\cite{Ang01b}, 
the divergence appears for those terms in the energy 
that originate from the interaction of a particle with 
its conjugated partner
\begin{eqnarray}
\label{eq:dft:div}
\mathcal{H}^{\text{EDF}} (\varphi)
& = & \bigg[
      \ldots 
      + \big(
           w^{\rho\rho}_{\mu\mu\mu\mu}
         + w^{\rho\rho}_{\bar\mu \bar\mu \bar\mu \bar\mu}
         + w^{\rho\rho}_{\bar\mu \mu \bar\mu \mu}
         + w^{\rho\rho}_{\mu \bar\mu \mu \bar\mu}
        \big)
          \frac{v^2_\mu e^{2i\varphi}}
               {u^2_\mu + v^2_\mu e^{2i\varphi}}
          \frac{v^2_\mu e^{2i\varphi}}
               {u^2_\mu + v^2_\mu e^{2i\varphi}}
       \nn \\
&   & \qquad
         + 4 w^{\kappa\kappa}_{\mu\bar{\mu}\mu\bar{\mu}}
           \frac{u_\mu v_\mu}
                {u^2_\mu + v^2_\mu e^{2i\varphi}}
           \frac{v_\mu v_\mu e^{2i\varphi}}
                {u^2_\mu + v^2_\mu e^{2i\varphi}}  
      + \ldots \;
      \bigg] \;  
      \mathcal{I} (\varphi)
.
\end{eqnarray}
For \mbox{$v_\mu^2 = 0.5$} and $\varphi = \pi/2$, the denominator
in the transition densities becomes zero. One of the two denominators 
is canceled by the same factor contained in $\mathcal{I} (\varphi)$, 
Eq.\ (\ref{eq:I:can}), while the other one causes the 
divergence\cite{Ang01b}.

2)
It can be shown from very general arguments\cite{Ben07a} that the 
Hamiltonian kernel should have a certain dependence on $\varphi$.
The divergent contributions to (\ref{eq:dft:div}) do not follow
this rule. This shows that such terms are spurious even for 
\mbox{$\varphi \neq \pi/2$}.

3) 
In the case of a two-body Hamiltonian the divergence disappears 
when one identifies $2 w^{\rho\rho}_{\mu \bar\mu \mu \bar\mu}
= 4 w^{\kappa\kappa}_{\mu \bar\mu \mu \bar\mu}
= \bar{v}_{\mu \bar\mu \mu \bar\mu}$ and
\mbox{$w^{\rho\rho}_{\mu\mu\mu\mu} = 0$}. This can be used 
to combine the $u_\mu$ and $v_\mu e^{2i\varphi}$ factors such that 
they cancel the dangerous denominator\cite{Ang01b}.

4)
The matrix elements of the kind $w^{\rho\rho}_{\mu\mu\mu\mu}$
might indeed have non-zero values in an EDF, which represent a spurious 
interaction of a particle with itself. They violate the exchange 
symmetry in a Fermionic systems (the Pauli principle), and lead to 
a spurious contribution to the total binding energy 
already on the mean-field level. The appearance of this so-called  
"self-interaction" is a well-known annoyance of EDFs for electronic 
systems\cite{Per81a}, but ignored in nuclear physics so far.

5) 
In the case of an EDF with pairing, there is an additional spurious 
"self-pairing" interaction that originates from the scattering of a pair 
onto itself. The interaction energy from two isolated Fermions 
occupying pair-conjugated orbitals divided by the occupation of the
pair should have the same value as if pairing correlations were not 
present\cite{Ben07a}. Again, this is violated by a general EDF. 
The actual expression for the spurious
self-pairing energy combines matrix elements and the occupation 
factors that weight them and will be given elsewhere\cite{Ben07a}.

6)
The divergent contributions to Eq.\ (\ref{eq:dft:div}) come from
terms that represent self-interaction and self-pairing. By that, 
the projection adds a second level of spuriosity to these terms.
Not the whole contribution from the self-interaction and self-pairing
terms is divergent, though. 

7)
The self-interaction and self-pairing  contribute to
the total energy not only because the relevant matrix elements have 
spurious non-zero values, but also because they are multiplied with 
unphysical weights. When evaluating the Hamiltonian kernel
(\ref{eq:Eproj:H}) by commutating the creation and annihilation 
operators of the Hamiltonian with those setting up the HFB states
until they hit the vacuum, it becomes clear that each pair of 
conjugated particles $(\mu, \bar\mu)$ can be multiplied with
one $e^{2i\varphi}$ factor stemming from the rotated state 
to the right only, and that the possible combinations of $u_\mu$ and 
$v_\mu$ for a given $\mu$ will always be bilinear. Only when 
the Wick theorem is applied, one obtains terms which are of 4th 
order in $u_\mu$ 
and $v_\mu$ and quadratic in $e^{2i\varphi}$. The reason is 
that multiple contributions from the same particle or pair of 
particles are not excluded from the sum as the Wick theorem implicitely 
assumes that the matrix elements these terms multiply are zero or
sum up to zero, so that they will not contribute anyway. In an EDF 
these matrix elements are not zero anymore, but the Wick theorem is 
still used.

One rigorous way to remove the spurious terms from an energy 
functional would be to remove all possible self-interactions
in the total energy by excluding that the same summation index
appears more than once in a contribution to Eq.\ (\ref{E:mf:dft:2})
This was, in fact, already used by Hartree in his seminal paper 
on the Hartree method\cite{Har28a}. Or, alternatively, one sums
up explicitely all the contributions where the same summation
index appears two times or more often and subtracts this as a 
self-energy correction from the total energy\cite{Per81a}. This leads, 
however, to enourmous complications in the variational equations,
particularly when extended to density-dependent interactions.
%
%--------------------------------------------------------------------------
%
\subsection{The origin of the finite step}
As pointed out by Stoitsov \etal\cite{Dob05a}, there is also a finite step 
that appears when passing the divergence as a function of a collective
coordinate. The analysis of its origin will be given elsewhere\cite{Ben07a}. 
Let us just outline the main arguments:
With the substitution \mbox{$z = e^{i \varphi}$}, the integral over the real 
gauge angle $\varphi$ in Eq.\ (\ref{eq:proj:number}) can be transformed 
into an integral in the complex plane, that can be analyzed with the
tools from function theory\cite{Dob05a,Ben07a}. In the Hamiltonian
case, the integral has a pole at \mbox{$z = 0$}, which has an order of 
the number of particles below the Fermi energy. The residue of this pole
is proportional to the projected energy. In a projected EDF 
framework, there are two
changes to this scenario: First, there is an additional contribution 
to the pole at \mbox{$z=0$}, this time from all particles, and second, 
there appear additional poles at $z_\mu^\pm = \pm i |u_\mu|/|v_\mu|$
along the imaginary axis, which also contribute for all particles 
below the Fermi energy. Both originate from the 
same terms as the divergence. The two additional contributions 
to the projected energy are huge (on the order of several hundered 
MeV for \nuc{18}{O}), but of opposite sign and nearly, but not exactly, 
canceling each other. The step appears when one of the poles at 
$\pm i |u_\mu|/|v_\mu|$ enters or leaves the integration contour 
\mbox{$|z|=1$} when the single-particle energy of the corresponding 
particle crosses the Fermi energy.
%
%==========================================================
%
\section{Further discussion}
Density-dependent terms add further complications to the divergence,
which we will not address here in detail. Let us just comment that, 
first, in a projected theory, the evaluation of a density-dependent 
term with non-integer power requires the evaluation of a (multivalued) 
root of a complex number, which leaves the Riemann sheet for choice.
Second, when one expands the
densities in the energy functional for density-dependent terms, and
combines the resulting terms similar to Eq.\ (\ref{eq:E:proj:dft:2}), 
one finds again terms which contain more than just one $v_{\mu} e^{2i\varphi}$ 
factor originating from $| \text{HFB}(\varphi) \rangle$ (one of them
with a usually non-integer power), which again introduces an unphysical
dependence on $\varphi$ into the Hamiltonian kernel.

There are good and profound reasons to use different effective interactions 
in the particle-hole and particle-particle channels. The particle-hole 
and particle-particle channel of the effective interaction sum up
different classes of diagrams\cite{Hen64a}, which inevitably leads 
to different expressions for the effective interaction in both channels.
Short-range correlations are resummed into the functional providing
the two channels with different density-dependent forms. On the other 
hand, the long-range correlations from large-amplitude fluctuations around 
the mean-field states are either neglected in a pure mean-field
approach when they are assumed to be small, or otherwise described 
explicitely by projection and GCM-type configuration mixing. 
A correction that removes the divergent part and the finite step 
from the projected energy can be set-up when identifying the terms 
with an unphysical dependence on the gauge angle comparing the 
expressions obtained from the standard and generalized Wick 
theorem\cite{Ben07a,Lac07a}.

While particle-number projection is the prominent example for the
appearance of a divergence, similar divergences can be expected 
in any GCM calculation or projection on any other quantum number. 
As in these cases the mixed states have different canonical bases,
the transition density matrix is usually not diagonal and
the analysis is less obvious. The widely used collective 
Schr{\"o}dinger equation and Bohr Hamiltonian that are set-up through 
a series of approximations (at some points including improvements) 
from exact configuration mixing cannot be expected to be free of 
problems stemming from the divergence either, although there they 
will be more difficult to identify.
%
%==========================================================
%
\section{Summary}
First, we have examined for the example of \nuc{120}{Sn} the effect 
of a minimization after particle-number projection and of GCM
ground state correlations from pairing vibrations. For \nuc{120}{Sn}, 
the effect is visible, but not enormous, and clearly smaller than 
the current uncertainties from the parameterization of the effective 
pairing interaction. This might be, however, different for other nuclei.

Second, we have examined the divergence that appears in 
particle-number projected EDF. As its origin we identify
the spurious self-pairing contribution contained in all common 
energy functionals for self-consistent mean-field models. 
On the mean-field level, the self-pairing as such adds a 
small spurious contribution to the total binding energy, that
is similar to, but smaller, than the usual spurious 
centre-of-mass or rotational energies from broken symmetries 
(in fact, together with the self-energy it is the spurious energy 
from violating the Pauli principle in approximate EDF approaches). 
When generalizing the EDF to particle number projection via the 
generalized Wick theorem, the self-pairing provides the Hamiltonian 
kernels with an unphysical dependence on the gauge angle, which 
ultimately leads to divergences and steps. 

%This raises profound questions about the nature, the proper
%definition, and unambiguous evaluation of projected EDF, which 
%have to be addressed, and hopefully satisfactorily answered,
%in the future.
%
%==========================================================
%
\section*{Acknowledgements}

The authors thank K.\ Bennaceur D. Lacroix, and P.-H.\ Heenen for 
many fruitful discussions on the theoretical and numerical treatment 
of dynamical pairing correlations, and J.\ Dobaczewski, W.\ Nazarewicz, 
M.\ V.\ Stoitsov, and  L.\ Robledo for many inspiring and clarifying 
discussions on the appearance and interpretation of, and the threat 
posed by, the divergence and step in particle-number projected EDF.
%
%==========================================================
%


\begin{thebibliography}{99}

\bibitem{RMP}
  M. Bender, P.-H. Heenen, and P.-G. Reinhard,
  Rev. Mod. Phys. \textbf{75}, 121 (2003).

\bibitem{Ben05b}
  M. Bender and P.-H. Heenen,
%  Proc. of ''The Fourth International Conference on Exotic Nuclei 
%  and Atomic Masses'' (ENAM'04),
%  Callaway Gardens, Pine Mountain, Georgia, USA, September 12-16, 2004.
%  C. J. Gross, W. Nazarewicz, K. P. Rykaczewski [edts.]
  Eur. Phys. J. A 25, s01 (2005) 519.

\bibitem{Man75a}
  H. J. Mang,
  Phys. Rep. \textbf{18}, 327 (1975).

\bibitem{Rin80a}
 P. Ring and P. Schuck,
 \emph{The Nuclear Many-Body Problem},
 Springer Verlag, New York, Heidelberg, Berlin, 438 (1980).

\bibitem{Dob84a}
  J. Dobaczewski, H. Flocard, J. Treiner,
  Nucl. Phys. \textbf{422}, 103 (1984).
 
\bibitem{SLyx}
  E. Chabanat, P. Bonche, P. Haensel, J. Meyer, and R. Schaeffer,
  Nucl. Phys. \textbf{A635}, 231    (1998);
  Nucl. Phys. \textbf{A643}, 441(E) (1998).

\bibitem{Rig99a}
  C. Rigollet, P. Bonche, H. Flocard, P.-H. Heenen,
  Phys. Rev. C \textbf{59}, 3120 (1999).

\bibitem{Gal94a}
  B. Gall, P. Bonche, J. Dobaczewski, H. Flocard, and P.-H. Heenen, 
  Z. Phys. \textbf{A348}, 183 (1994).

\bibitem{Flo97a}
  H. Flocard and N. Onishi,
  Ann. Phys. \textbf{254} (1997) 275.

\bibitem{She00a}
  J. A. Sheikh and P. Ring,
  Nucl. Phys. \textbf{A665}, 71 (2000).

\bibitem{Sto06a}
  M. V. Stoitsov, J. Dobaczewski, R. Kirchner, W. Nazarewicz, J. Terasaki,
  preprint nucl-th/0610061.

\bibitem{Hee93a}
  P.-H. Heenen, P. Bonche, J. Dobaczewski, H. Flocard,
  Nucl. Phys. \textbf{A561}, 367 (1993).

\bibitem{Fom70a}
  V. N. Fomenko,
  J. Phys. (G.B) A \textbf{3}, 8 (1970).

\bibitem{Hil53a}
  D. L. Hill and J. A. Wheeler, 
  Phys. Rev. \textbf{89}, 1106 (1953);\\
  J. J. Griffin and J. A. Wheeler,
  Phys. Rev. \textbf{108}, 311 (1957).

\bibitem{Bon90a}
  P. Bonche, J. Dobaczewski, H. Flocard, P.-H. Heenen and J. Meyer, 
  Nucl. Phys. \textbf{A510}, 466 (1990).

\bibitem{Oni66a}
  N. Onishi and S. Yoshida,
  Nucl. Phys. \textbf{80}, 367 (1966).

\bibitem{Bal69a}
  R. Balian and E. Br{\'e}zin, 
  Il Nuovo Cimento, B \textbf{64}, 37 (1969).

\bibitem{Bay86a}
  D. Baye and P.-H. Heenen,
  J. Phys. \textbf{A19}, 2041 (1986).

\bibitem{Val00a}
  A. Valor, P.-H. Heenen, P. Bonche, 
  Nucl. Phys. \textbf{A671}, 145 (2000).

\bibitem{Ben03a}
  M. Bender and P.-H. Heenen, 
  Nucl. Phys. \textbf{A713}, 390 (2003).

\bibitem{Ben03b}
  M. Bender, H. Flocard, P.-H. Heenen, 
  Phys. Rev. C \textbf{68}, 044321 (2003).

\bibitem{Dug03a}
  T. Duguet, M. Bender, P. Bonche, P.-H. Heenen, 
  Phys. Lett. \textbf{B559}, 201 (2003).

\bibitem{Ben04b}
  M. Bender, P. Bonche, T. Duguet, P.-H. Heenen, 
  Phys. Rev. C \textbf{69}, 064303 (2004).

\bibitem{Ben04c}
  M. Bender, P.-H. Heenen, P. Bonche, 
  Phys. Rev. C \textbf{70}, 054304 (2004).

\bibitem{Ben05a}
  M. Bender, G. F. Bertsch, P.-H. Heenen,
  Phys. Rev. Lett. \textbf{94}, 102503 (2005);\\
  Phys. Rev. C \textbf{73}, 034322 (2006).

\bibitem{Ben06a}
  M. Bender, P. Bonche, P.-H. Heenen,
  Phys. Rev. C \textbf{74}, 024312 (2006).

\bibitem{Madrid}
  R. R. Rodriguez-Guzman, J. L. Egido, and L. M. Robledo, 
  Phys. Rev. C \textbf{62}, 054319 (2002);
  J. L. Egido, L.M. Robledo, 
%  in 
%  \emph{Extended Density Functionals in Nuclear Physics}, 
%  G. A. Lalazissis, P. Ring, D. Vretenar [edts.], 
  Lecture Notes in Physics No. 641 (Springer, Berlin, 2004), 
  p. 269.

\bibitem{Nik06a}
  T. Nik{\v s}i{\'c}, D. Vretenar, and P. Ring,
  Phys. Rev. C \textbf{73}, 034308 (2006).

%\bibitem{Schm72a}
%  H. Schmidt,
%  in ``Developments and Borderlines of Nuclear Physics'',
%  Proceedings of the International School of Physics ``Enrico Fermi'',
%  course LIII, Varenna, Italy, 19-31 July 1971.
%  Academic Press, New York and London, 1972, page 144.

\bibitem{Rip69a}
  G. Ripka and R. Padjen,
  Nucl. Phys. \textbf{A132}, 489 (1969).

%\bibitem{Jus69a}
%  D. Justin, M. V. Mihailovic and M. Rosina,
%  Phys. Lett. \textbf{B29}, 458 (1969);
%  Nucl. Phys. \textbf{A182}, 54 (1972).

%\bibitem{Sie72a}
%  C. D. Siegal and R. A. Sorensen,
%  Nucl. Phys. \textbf{A184}, 81 (1972).

\bibitem{Fae73a}
%  A. Faessler, F. Gr{\"u}mmer, A. Plastino,
%  Z. Phys. \textbf{260}, 305 (1973);\\
  A. Faessler F. Gr{\"u}mmer, A. Plastino, F. Krmpotic,
  Nucl. Phys. \textbf{A217}, 420 (1973).

\bibitem{Mey91a}
  J. Meyer, P. Bonche, J. Dobaczewski, H. Flocard, P.-H. Heenen,
  Nucl. Phys. \textbf{A533}, 307 (1991).

\bibitem{Hee01a}
  P.-H. Heenen, A. Valor, M. Bender, P. Bonche, H. Flocard
  Eur. Phys. J. \textbf{A11}, 393 (2001) .

\bibitem{pairconstraint}
  M. Bender, K. Bennaceur, and T. Duguet,
  unpublished.

\bibitem{Don98a}
  F. D{\"o}nau,
  Phys. Rev. C \textbf{58}, 872 (1998).

\bibitem{Ang01b}
  M. Anguiano, J. L. Egido, and L. M. Robledo,
  Nucl. Phys. \textbf{A696}, 467 (2001).

\bibitem{Dob05a}
  J. Dobaczewski, W. Nazarewicz, P.-G. Reinhard, and M. V. Stoitsov,
  in preparation.

\bibitem{Ben07a}
  M. Bender and T. Duguet,
  in preparation.

\bibitem{Koh64a}
  W. Kohn and L. J. Sham,
  Phys. Rev. \textbf{137}, A1697 (1964);\\
  W. Kohn, 
  Rev. Mod. Phys. \textbf{71}, 1253 (1998).

\bibitem{Oli88a}
  L. N. Oliveira, E. K. U. Gross, and W. Kohn,
  Phys. Rev. Lett. \textbf{60}, 2430 (1988).

\bibitem{Per81a}
  J. P. Perdew and A. Zunger,
  Phys. Rev. \textbf{B23}, 5048 (1981).

\bibitem{Har28a}
  D. R. Hartree,
  Proc. Cambridge Philos. Soc., Vol. 1928, 89 (1928).

\bibitem{Lac07a}
  D. Lacroix, T. Duguet and M. Bender ,
  in preparation.

\bibitem{Hen64a}
  E. M. Henley and L. Wilets,
  Phys. Rev. \textbf{133}, B 1118 (1964).

\end{thebibliography}
\end{document}